\documentclass{article}

\usepackage{microtype}
\usepackage{graphicx}
\usepackage{multirow} 
\usepackage{subfigure}
\usepackage{booktabs} % for professional tables

\usepackage{hyperref}

\usepackage[accepted]{icml2025} 
\makeatletter
\providecommand{\icml@copyright}{} % 如果没定义，就定义为空
\makeatother
\usepackage{amsmath}
\usepackage{amssymb}
\usepackage{mathtools}
\usepackage{amsthm}

\usepackage[capitalize,noabbrev]{cleveref}

\theoremstyle{plain}

\theoremstyle{definition}

\theoremstyle{remark}

\usepackage[textsize=tiny]{todonotes}

\begin{document}

\twocolumn[
\icmltitle{Breaking the Cold-Start Barrier: Reinforcement Learning with Double and Dueling DQNs}



\icmlsetsymbol{equal}{*}

\begin{icmlauthorlist}
  \icmlauthor{Minda Zhao\thanks{Correspondence to: Minda Zhao <mindazhao@hsph.harvard.edu>.}}{}
\end{icmlauthorlist}


\vskip 0.3in
]

\footnotetext{Correspondence: mindazhao@hsph.harvard.edu. Harvard University, Boston, United States.}




\begin{abstract}

Recommender systems struggle to provide accurate suggestions to new users with limited interaction history, a challenge known as the cold-user problem. This paper proposes a reinforcement learning approach using Double and Dueling Deep Q-Networks (DQN) to dynamically learn user preferences from sparse feedback, enhancing recommendation accuracy without relying on sensitive demographic data. By integrating these advanced DQN variants with a matrix factorization model, we achieve superior performance on a large e-commerce dataset compared to traditional methods like popularity-based and active learning strategies. Experimental results show that our method, particularly Dueling DQN, reduces Root Mean Square Error (RMSE) for cold users, offering an effective solution for privacy-constrained environments.

\end{abstract}

\section{Introduction}
\label{submission}

Online platforms like Amazon, Youtube, Spotify, and Tiktok rely on recommender systems to help users navigate an overwhelming number of options. As product variety grows, users face choice overload, where being confronted with too many possibilities can reduce satisfaction \cite{schwartz2004}. Recommenders mitigate this by personalizing content to each user’s tastes. Nevertheless, providing useful recommendations for new users – the cold-user problem – remains a significant challenge. Cold users come with little or no interaction history, so collaborative filtering and matrix factorization models have sparse or zero data to learn from, often resorting to generic “popular item” suggestions\cite{palomares2018}. This can degrade the user experience for newcomers. The problem is exacerbated by privacy regulations like General Data Protection Regulation (GDPR) , which restrict using demographic \cite{pazzani1999} or social \cite{golbeck2006} media data without consent, limiting the available information about new users.

\begin{figure}[ht]
\vskip 0.2in
\begin{center}
\centerline{\includegraphics[width=1.1\columnwidth]{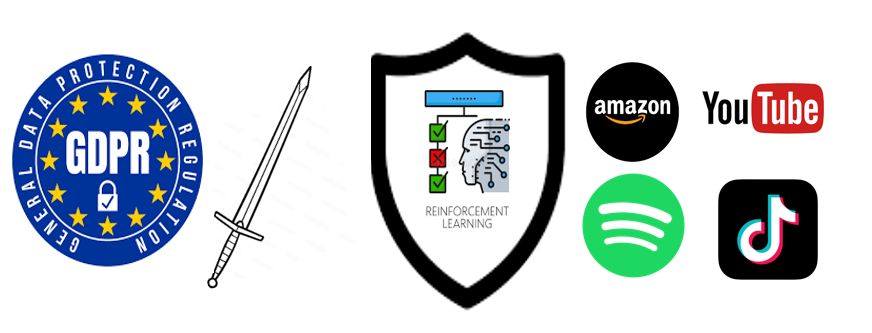}}
\caption{Illustration of the GDPR-compliant cold-start recommendation framework: GDPR privacy constraints (left) limit access to demographic/social data, a reinforcement-learning agent (center) infers new-user preferences from sparse interactions, and the approach targets large-scale platforms (Amazon, YouTube, Spotify, TikTok).}
\label{icml-historical}
\end{center}
\vskip -0.2in
\end{figure}

Prior solutions to the cold-user problem have included asking users to provide initial feedback or leveraging side information. For example, active learning approaches solicit a new user’s preferences by prompting them to rate a few items, thereby quickly gathering personal data \cite{elahi2016}. Such methods can improve cold-start recommendations but depend on users’ willingness to give explicit feedback and are constrained by which items are queried. Content-based methods incorporate user attributes or item descriptions to infer preferences for new users \cite{Zhou2011}. For instance, functional matrix factorization uses user profile features or product content to bootstrap recommendations \cite{Zhou2011}. While these approaches alleviate cold-start issues, they require additional data sources or interaction steps. 

Reinforcement learning (RL) has emerged as a promising paradigm for recommender systems, offering a way to continuously adapt to user interactions and optimize long-term rewards \cite{wang2014}. Unlike one-shot recommendation algorithms, an RL agent can learn an interactive policy: it treats recommendation as a sequential decision process, where at each step the agent recommends an item, observes the user’s reaction (click, purchase, etc.), and updates its strategy \cite{zhao2018}. This framework can naturally handle cold users by learning how to explore their preferences in real time, without requiring demographic data or upfront ratings. However, using RL for recommender systems is rare bacause for the action space, the set of possible items to recommend, is typically huge, making learning difficult \cite{dureddy2018}. 

Recently, Giannikis et al. \citeyear{GIANNIKIS2024111752} introduced a novel approach that combines reinforcement learning with a model-based recommender to tackle the cold-user problem. In their framework, an agent uses a Deep Q-Network (DQN) to choose which item to show to a new user at each step. They demonstrated that an item-based DQN policy could outperform traditional strategies like popularity or hand-crafted active learning heuristics in terms of rating prediction accuracy. This approach illustrates the potential of deep reinforcement learning for cold-start recommendation, but also leaves room for improvement. Standard DQN algorithms are known to suffer from issues such as overestimation bias in Q-value predictions \cite{vanhasselt2015deepreinforcementlearningdouble}, which can lead to suboptimal policies, and they might not fully capitalize on the underlying structure of the state–action value function.

In this paper, we build upon the baseline of Giannikis et al. \citeyear{GIANNIKIS2024111752} by introducing enhanced DQN architectures to improve cold-user recommendations. In particular, we investigate two DQN variants: Double DQN \cite{vanhasselt2015deepreinforcementlearningdouble} and Dueling DQN \cite{wang2016duelingnetworkarchitecturesdeep}. Double DQN addresses the overestimation problem by decoupling action selection from evaluation, using two Q-networks to obtain more accurate value estimates. Dueling DQN features a neural network architecture with separate streams to estimate state values and action advantages, which can lead to more stable and generalized learning of which states are valuable regardless of action \cite{wang2016duelingnetworkarchitecturesdeep}. These techniques have proven effective in other domains, but have not been explored in recommender system contexts. We adapt them to the cold-user recommendation scenario and show that they yield small yet consistent gains in accuracy.

Our contributions are as follows. (1) We formulate the cold-user recommendation task as an RL problem and reproduce a state-of-the-art baseline DQN method augmented with matrix factorization for new users. (2) We propose the use of Double and Dueling DQN architectures in this setting and describe how they can alleviate specific deficiencies of the standard DQN approach (like overestimation bias and slow convergence). (3) We conduct extensive experiments on a large real-world retail dataset, evaluating recommendation performance across various display sizes (number of items shown) and comparing against multiple baseline strategies (non-personalized and active learning-based). (4) We demonstrate that our enhanced RL agents achieve the lowest Root Mean Square Error (RMSE) in predicting user interactions for cold users – outperforming the original DQN and other baselines – and analyze the statistical significance of the improvements. (5) We discuss the implications of deploying RL for cold-start users, the limitations observed (e.g. performance trends as more items are recommended), and suggest future directions such as reward shaping, hybrid methods, and incorporating richer success metrics beyond RMSE.

The rest of this paper is organized as follows. Section 2 reviews related work on recommender systems, cold-start problems, and reinforcement learning in recommendations. Section 3 formally states the cold-user recommendation problem we address. Section 4 describes our methodology, including the baseline matrix factorization + DQN approach and the proposed Double and Dueling DQN enhancements, as well as details of the dataset and evaluation metrics. Section 5 outlines the experimental setup, model architecture, and Section 6. Section 7 presents our findings, discusses the limitations of the proposed approach, outlines avenues for future work, evaluates its feasibility, and offers concluding remarks.

\section{Related Work}
\label{submission}

\subsection{Cold-User Problem}

The cold-user problem is well studied because it is one of the most popular topic in recommender systems today especially for the internet shop \cite{ricci2015}. Traditional collaborative filtering algorithms, including neighborhood methods and latent factor models, perform poorly for new users due to lack of past interactions \cite{iyengar2001}. To combat this, various strategies have been developed. One category is content-based and hybrid methods that leverage side information about users or items. For example, Zhou et al. \cite{Zhou2011} propose functional matrix factorizations that incorporate user attributes (e.g. demographics or survey responses) into the factor model to make initial predictions for new users. Such methods can partially alleviate cold-start by using proxy data, but obtaining rich user attributes is not always feasible (especially under privacy constraints) and may not capture individual taste nuances.

Another line of work uses interview and active learning techniques. These approaches proactively acquire information from the user \cite{elahi2016}. Based on those responses, a more personalized model is trained. Prior research by Geurts and Frasincar \citeyear{geurts2017} explored active learning heuristics for cold-user recommendation. They introduced strategies like Gini (which picks items with the highest information gain as queries) and Pop-Gini (a hybrid that balances popularity with information gain) to decide which items to present to a new user for feedback. Such strategies help profile new users more quickly than waiting for organic interactions. However, they require explicit feedback and interrupt the user experience with questions, which not all users may engage with. Moreover, these heuristics need to be carefully designed to select informative items; if the user’s tastes lie outside the queried items, the value of the feedback is limited \cite{smith2017}.

\subsection{Reinforcement Learning in Recommender System}

In recent years, reinforcement learning has been applied to personalization problems to optimize long-term user satisfaction or cumulative reward in recommender system. RL can naturally handle interactive settings like news recommendation, ad placement, or multimedia recommendation, where each user interaction (click/no-click, watch or skip) provides a reward signal. Wang et al. \citeyear{wang2014} presented an RL approach for music recommendation that used an $\varepsilon$-greedy strategy to balance exploring new artists with exploiting known preferences. The RL paradigm is appealing for recommendations because it can in principle adjust to user reactions on the fly and even learn an optimal strategy for asking questions or choosing content for cold users. Multi-armed bandits (a stateless form of RL) have also been widely studied for recommendation and ads, focusing on the exploration-exploitation trade-off to maximize clicks or sales \cite{tang2019, lin2022}. However, full RL (with state) can consider the sequence of recommendations and dependencies across time, which bandits ignore \cite{huang2021}.

One challenge in applying RL to recommender systems is the enormous action space – there may be hundreds of thousands or millions of items to choose from at each step. Traditional RL algorithms struggle with such large discrete action spaces because it’s infeasible to try every action enough times, and function approximation (like deep neural networks) becomes necessary to generalize across actions. Dureddy and Kaden \cite{dureddy2018} highlighted this issue in their work on RL for collaborative filtering, noting that naive RL approaches can perform poorly due to the huge item space and sparse rewards. Approaches to mitigate this include action space reduction \cite{choi2018} or using parametric policies that rank items rather than evaluate each item independently \cite{9514509}.

\subsection{Reinforcement Learning for Cold Users Problem}

Combining RL with cold-start scenarios has begun to garner attention. The idea is to let an agent learn an optimal strategy to interact with a new user, such as which items or questions to present, in order to quickly identify their preferences and maximize some notion of cumulative reward (like total number of liked items) \cite{dureddy2018}. The baseline for our work is the method by Giannikis et al. \cite{GIANNIKIS2024111752}, who formulated cold-user recommendation as an RL problem. In their framework, an episode corresponds to a new user session. At each time step, the agent chooses an item to show from a restricted pool, and the user’s reaction is observed implicitly. A pre-trained matrix factorization model is used to estimate user-item interaction probabilities, which effectively provides a prior for the agent. They investigated two variants: an item-based RL approach where the agent’s actions are individual items \cite{smith2017}, and an AL-based RL approach \cite{dureddy2018} where actions correspond to selecting one of the predefined active learning strategies (like PopGini) to apply. In their experiments on a large retail dataset, the item-based DQN approach achieved an average RMSE of 0.430 in predicting user interactions. Their RL agent outperformed other baselines such as pure popularity, entropy-based querying, etc., particularly when the number of recommended items was moderate (25 or 50). However, they observed that for very short lists (10 items), a trivial random strategy achieved a surprisingly low RMSE (around 0.378) – a result of the metric and data sparsity (if a user interacts with none of the randomly shown items, the error is low) – and for long lists (100 items), simpler strategies caught up with or surpassed the RL approach. These observations indicate that while RL is promising for cold-start, there is room to improve its robustness and consistency across scenarios.

The success of Deep Q-Networks (DQN) in game-playing and other domains has led to several extensions that improve learning stability and performance. Double DQN was introduced by van Hasselt et al. \cite{vanhasselt2015deepreinforcementlearningdouble} to address overestimation of action values in DQN. Standard DQN uses one network to both select and evaluate an action, which tends to overestimate Q-values due to maximization bias. Double DQN instead uses two networks (often the same network at different target intervals): one is used to pick the best action, and the other (a target network) is used to evaluate that action’s value, resulting in more accurate Q-value updates. This method has been shown to reduce overestimation and often improves final policy performance. Dueling DQN, proposed by Wang et al. \cite{wang2016duelingnetworkarchitecturesdeep}, changes the network architecture by splitting the Q-value computation into two parts: one stream estimates the state value (how good the state is overall), and another estimates the advantage of each action (how much better an action is compared to the average for that state). These two are combined to form the final Q-values. The dueling architecture can learn state values even when the choice of action doesn’t matter (which can happen in many states), and it can more easily identify salient actions in states where actions do make a difference \cite{doi:10.1177/03611981231205877}. This often speeds up learning and leads to more robust policies, as the network can generalize the value of similar states without being confused by irrelevant action specifics. Both Double and Dueling DQN have been successfully employed in various environments (often combined with other improvements like prioritized replay \cite{schaul2016prioritizedexperiencereplay}, as in the Rainbow algorithm \cite{hessel2018}), but they have not been widely applied in recommendation tasks yet. In the context of cold-user recommendation, we hypothesize that Double DQN can yield more reliable estimates of the long-term reward for recommending certain items, preventing the agent from over-valuing suboptimal items due to noise, while Dueling DQN can help distinguish between states (users with different emerging preference patterns) and actions (items) more effectively even from sparse feedback.

Our work contributes to the intersection of RL and recommender systems by bringing these advanced DQN techniques into the cold-start recommendation problem. To our knowledge, this is one of the first studies applying Double and Dueling network architectures to recommender system scenarios. We compare these against the baseline single-network DQN approach and traditional strategies, providing insight into how each fares in terms of prediction accuracy for new users.

\section{Problem Statement}
\label{submission}

We focus on the cold-user recommendation problem: how to recommend items to a new user who has no prior interaction history in the system \cite{ricci2015}. Formally, let $U$ be the set of users and $I$ the set of items. A new user $u \in U$ has no observed interactions (purchases, ratings, clicks, etc.) in the training data. We want to generate a list of $k$ items $[i_1, i_2, \dots, i_k]$ to present to user $u$ such that the user will engage positively with those items. A positive engagement could be a purchase or a high rating, while a negative outcome could be ignoring the item or returning a purchased item. The fundamental challenge is that we do not know $u$’s preferences beforehand – any inference about $u$ must be drawn either from generic patterns (like popularity or item similarity) or from interactions with the user in real time. Our objective is to maximize the utility of the recommendations for the cold user, which could be defined in various ways (click-through rate, purchase rate, total value of items purchased, etc.). In this work, we focus on predictive accuracy of user-item interactions as a measurable proxy: essentially, we want the recommender’s predictions of the user’s behavior to be as close as possible to the user’s actual behavior for the recommended items.

Our goal is to learn an optimal recommendation policy $\pi^*$ for cold users, which selects items sequentially based on the current state $s_t$. Since no prior data is available, the policy must infer preferences from session interactions, which we model using model-free deep reinforcement learning (Q-learning). Specifically, we investigate whether (1) RL agents can outperform heuristic strategies, (2) Double and Dueling DQNs improve accuracy and stability over standard DQN, and (3) these gains are consistent across different recommendation list lengths.

\section{Method}
\label{submission}

\begin{figure*}[h]
    \centering
    \includegraphics[width=0.7\textwidth]{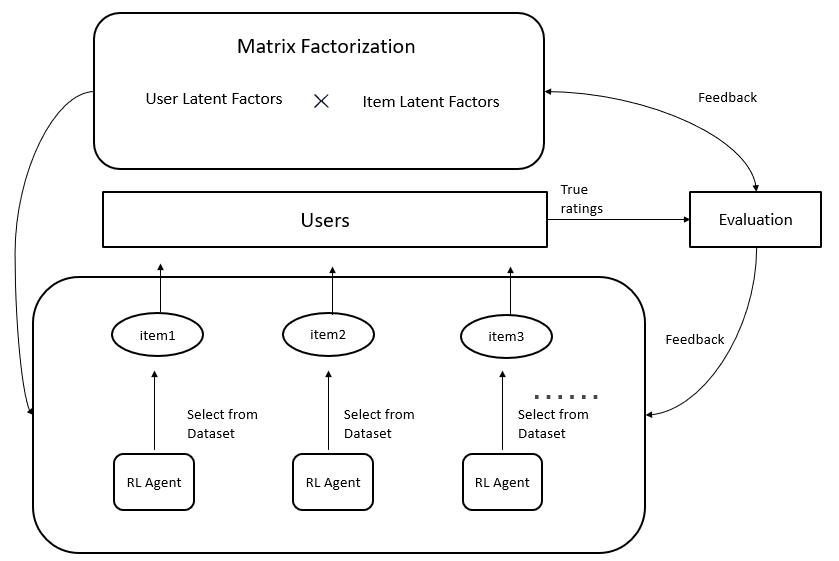}
    \caption{Overview of the models' architectures and interaction.}
    \label{fig:overview}
\end{figure*}

Our method consists of two main components: a Matrix Factorization (MF) model that provides a baseline recommendation score for any user-item pair, and a Deep Q-Network (DQN) agent that uses those scores and observed feedback to decide which item to recommend next to a cold user. We provide the overview of the whole structure of the model as in Figure \ref{fig:overview}.

\subsection{Matrix Factorization}

We train a standard matrix factorization model on historical interaction data to capture general preference patterns. MF is a popular collaborative filtering technique that represents each user $u$ and item $i$ in a latent factor space \cite{koren2009}. It learns user vectors $p_u \in \mathbb{R}^d$ and item vectors $q_i \in \mathbb{R}^d$ such that the dot product $p_u \cdot q_i$ (plus possibly biases) approximates the interaction $r_{ui}$ (e.g. a rating or implicit feedback signal). In our implicit feedback setting, we treat purchases and returns as signals. One simple approach is to assign a target value (like $r_{ui}=1$ for a purchase and $r_{ui}=-1$ for a return, or 0 for no interaction) and train MF to predict these. We optimize the MF model using an appropriate loss (e.g. regularized squared error or logistic loss if interpreting scores probabilistically). After training, for any user-item pair, MF can produce a predicted score $\hat{r}_{ui}^{MF} = p_u \cdot q_i$. 

\subsection{Active Learning strategies}

We employ nine non‐personalized AL strategies to rank items for a cold user to rate.  Table~\ref{tab:al_strategies} lists each strategy name and its scoring formula.  After computing $s(i)$ for all items, we query the user to rate the top-$k$ items in the chosen ranking.

\begin{table}[ht]
\centering
\caption{Non‐personalized AL strategies for cold‐start recommendation}
\label{tab:al_strategies}
\begin{tabular}{ll}
\toprule
\textbf{Strategy} & \textbf{Score $s(i)$} \\
\midrule
Popularity   & $\mathrm{pop}(i)$ \\
Entropy      & $-\sum_{j\in\{0,1\}}p(j\!\mid\!i)\,\log_{2}p(j\!\mid\!i)$ \\
Gini         & $1-\sum_{j\in\{0,1\}}[p(j\!\mid\!i)]^{2}$ \\
Variance     & $\tfrac{1}{|U_i|}\sum_{u\in U_i}(r_{ui}-\bar r_i)^{2}$ \\
Error        & $1-\max\{p(0\!\mid\!i),\,p(1\!\mid\!i)\}$ \\
PopEnt       & $w_{1}\log\mathrm{pop}(i)\;+\;w_{2}\,\mathrm{Entropy}(i)$ \\
PopGini      & $w_{1}\log\mathrm{pop}(i)\;+\;w_{2}\,\mathrm{Gini}(i)$    \\
PopVar       & $w_{1}\log\mathrm{pop}(i)\;+\;w_{2}\,\mathrm{Variance}(i)$\\
PopError     & $w_{1}\log\mathrm{pop}(i)\;+\;w_{2}\,\mathrm{Error}(i)$   \\
\bottomrule
\end{tabular}
\end{table}

In the table above, the five single‐heuristic strategies each target a specific criterion:  
\textit{Popularity} selects items with the most interactions (attention‐based);  
\textit{Entropy}, \textit{Gini}, and \textit{Variance} choose items with the most diverse feedback (uncertainty‐based);  
\textit{Error} picks items where the majority vote is least decisive (error‐reduction).  
The four combined‐heuristic strategies mix a log‐scaled popularity term with each corresponding single score (Entropy, Gini, Variance, or Error) via weights $w_{1},w_{2}$, balancing broad appeal and informativeness in item selection.

\subsection{Deep Reinforcement Learning}

\subsubsection{Deep Q-Network (DQN)}

We model the cold-user recommendation problem as a reinforcement learning task and employ a Deep Q-Network (DQN) to learn the action-value function. The DQN uses a neural network $Q(s,a;\theta)$ to approximate the expected cumulative reward of taking action $a$ (e.g. recommending an item) in state $s$. Training is done by minimizing the temporal-difference error between the predicted $Q$-value and a target $y$ derived from the Bellman equation. In particular, DQN optimizes the loss:

\begin{equation}
L(\theta) \;=\; \mathbb{E}_{(s,a,r,s')\sim \mathcal{D}}\Big[\,\big(r + \gamma \max_{a'} Q(s', a';\,\theta^{-}) \;-\; Q(s, a;\,\theta)\big)^2\Big]\,
\end{equation}

where $r$ is the reward, $\gamma$ is the discount factor, and $\theta^{-}$ are the parameters of a target network (a periodic copy of the Q-network used to compute a stable target). By using experience replay and a fixed target network, DQN stabilizes training and can learn an effective policy for cold-start recommendations.

\subsubsection{Double DQN}

While DQN achieved success in many domains, it is known to overestimate Q-values because the same values are used to both select and evaluate actions in the target. We address this by adopting Double DQN, which decouples action selection from evaluation to reduce overestimation bias. Instead of using $\max\_{a'}Q(s',a';\theta^-)$ for the target, Double DQN uses the online network $theta$ to choose the best action and the target network $\theta$ to evaluate its value. Formally, the target value in Double DQN is:

\begin{equation}
y^{\text{Double}} \;=\; r \,+\, \gamma \, Q\!\Big(s',\; \arg\max_{a'}Q(s',a';\,\theta)\,;\;\theta^{-}\Big)\, 
\end{equation}

This way, the action $a'$ that maximizes $Q(s',a';\theta)$ is picked (greedy selection by the current network), but its Q-value is estimated using the older target network. By decoupling these roles, Double DQN provides more accurate target estimates, leading to more reliable Q-learning updates.

\vspace*{3em}
\subsubsection{Dueling DQN}

We further integrate the Dueling DQN architecture to improve value estimation. The dueling network factorizes the Q-value into two separate estimators: a state-value function $V(s)$, which estimates the overall value of state $s$, and an advantage function $A(s,a)$, which captures how much better action $a$ is compared to the average action at state $s$. These two streams are combined to output $Q(s,a)$ as:

\begin{equation}
Q(s,a) \;=\; V(s) \;+\; A(s,a)\;-\; \frac{1}{|\mathcal{A}|}\sum_{a'}A(s,a')\,
\end{equation}

i.e. the advantage is normalized by subtracting its mean over actions. This decomposition ensures the advantages $A(s,a)$ sum to zero (making $V$ identifiable) while preserving $Q(s,a)=V(s)$ for the best action. Intuitively, the dueling architecture enables the agent to learn **state values even when actions have similar outcomes, and to learn advantages that differentiate actions when needed. This is especially useful in cold-start recommendation scenarios with many items of comparable relevance, as the network can more effectively learn which states (user contexts) are valuable and which actions yield above-average benefit in each state.

\subsection{evaluation metric}

We use Root Mean Square Error (RMSE) to measure how well our recommendation strategies predict user interactions for new (cold) users. For each test user and strategy—whether it’s our reinforcement learning (RL) agents or baseline methods—we simulate recommending a set number of items, such as 10, 25, 50, or 100. 

The RMSE is then calculated with the formula:
\begin{equation}
\text{RMSE} = \sqrt{\frac{1}{n} \sum_{(u,i) \in T} (\hat{r}_{ui} - r_{ui})^2}
\end{equation}

where:
\begin{itemize}
    \item \( r_{ui} \) is the true interaction of user \( u \) for item \( i \), defined as:
    \[
    r_{ui} = \begin{cases} 
    1 & \text{if the user purchases the item}, \\ 
    -1 & \text{if the user returns the item}, \\ 
    0 & \text{if there is no interaction},
    \end{cases}
    \]
    \item \( \hat{r}_{ui} \) is the predicted interaction.
\end{itemize}

The RMSE metric is used to evaluate the performance of the matrix factorization model. For the RL model, we adapt this evaluation framework through a distinct reward mechanism that aligns with RL's objective of cumulative reward maximization. Specifically, the RL agent interacts with an environment where its recommendations generate validation-based RMSE scores calculated on the data set separate from the matrix factorization training data. The reward signal for the RL agent is then defined as the reciprocal of these environment-generated RMSE values:

\begin{equation}
\text{Reward} = \frac{1}{\text{RMSE}}
\end{equation}

This design ensures the RL agent optimizes for recommendation quality through indirect RMSE minimization while maintaining proper separation between the matrix factorization model's training data and the RL agent's decision-making process.

\section{Experiment}

\subsection{Dataset and Preprocessing}

We evaluate our item-based DQN approach on a real-world e-commerce dataset provided by De Bijenkorf, a Dutch online retailer, which was used in prior research on cold-start recommendation \cite{10.1145/3106426.3106431}. Its implicit feedback dataset, which contains 2,563,878 binary interactions (purchase = 1, return = 0) from 563,495 users and 242,020 items over one year \cite{GIANNIKIS2024111752}. To simulate cold-start, we randomly select 25\% of users as \emph{cold users} and hide all their interactions from the training set.The remaining 75\% of users (warm users), together with cold-user feedback revealed during DQN interactions, train the matrix factorization (MF) model. Cold users with no initial feedback are excluded, and for the retained ones, their unseen interactions serve as the test set for final RMSE evaluation.
\subsection{Item-based Reinforcement Learning Environment}

 We model the cold-start interview as a Markov decision process in which, at each step, the agent selects one of the 200 most popular items (by frequency in the training set) to recommend. The state is represented by a 200-dimensional binary vector that marks which of these popular items have already been shown to the user. When the agent chooses action \(a_t\), the environment reveals whether the user interacted with item \(a_t\), and the corresponding entry in the state vector flips from 0 to 1.

\subsection{Model Configuration and Training}
Our MF backbone learns latent factors $p_u,q_i\in\mathbb{R}^{10}$ by minimizing squared error with learning rate 0.001 and $\ell_2$ regularization 0.01 over 100 iterations.  

The DQN agent implements a feed‐forward network with two hidden layers of 64 and 32 units (tanh activations) and uses the Huber loss. We train three variants—standard DQN \cite{mnih2013}, Double DQN \cite{vanhasselt2015deepreinforcementlearningdouble}, and Dueling DQN \cite{wang2016duelingnetworkarchitecturesdeep}—under identical settings via the Stable Baselines framework. Each agent interacts for 2,000 episodes per list size. 

\begin{table}[h]
  \centering
  \caption{Hyperparameters for MF and Item-based DQN}
  \label{tab:hyperparams}
  \begin{tabular}{lll}
    \toprule
    \textbf{Component} & \textbf{Parameter}         & \textbf{Value}           \\
    \midrule
    \multirow{4}{*}{MF model} &
      latent factors            & 10                        \\
    & learning rate              & 0.001                     \\
    & regularization             & 0.01                      \\
    & iterations                 & 100                       \\
    \midrule
    \multirow{9}{*}{DQN agent} &
      discount factor $\gamma$   & 0.99                      \\
    & learning rate               & $4\times10^{-4}$          \\
    & hidden layers               & [64, 32]                  \\
    & activation                  & tanh                      \\
    & buffer size                 & 100                       \\
    & batch size                  & 32                        \\
    & episodes                    & 2000                      \\
    & $\epsilon_0\to\epsilon_{\min}$ & 1.0 $\to$ 0.01        \\
    & target update freq          & 100 steps                 \\
    & loss function               & Huber                     \\
    \bottomrule
  \end{tabular}
\end{table}

\section{Result}

\begin{table}[htbp]
  \centering
  \caption{Comparison of strategies across different numbers of shown items (RMSE)}
  \setlength{\tabcolsep}{3pt}
  \label{tab:strategy_rmse}
  \begin{tabular}{lccccc}
    \toprule
    & \multicolumn{4}{c}{\textbf{Shown items}} & \\[-0.4em]
    \cmidrule(lr){2-5}
    \textbf{Strategy} & \textbf{10} & \textbf{25} & \textbf{50} & \textbf{100} & \textbf{Mean} \\
    \midrule
    Double~DQN          & 0.425 & \textbf{0.426} & 0.439 & 0.429 & 0.430 \\
    Dueling~DQN         & \textbf{0.408} & 0.436 & 0.432 & 0.441 & \textbf{0.429} \\
    Standard DQN        & 0.431 & 0.428 & \textbf{0.421} & 0.442 & 0.430 \\
    Popularity strategy & 0.497 & 0.471 & 0.469 & 0.438 & 0.469 \\
    Gini strategy       & 0.672 & 0.594 & 0.583 & 0.540 & 0.597 \\
    Entropy strategy    & 0.628 & 0.515 & 0.583 & 0.550 & 0.569 \\
    Error strategy      & 0.513 & 0.594 & 0.553 & 0.585 & 0.561 \\
    Variance strategy   & 0.559 & 0.482 & 0.554 & 0.580 & 0.544 \\
    PopGini strategy    & 0.466 & 0.480 & 0.458 & 0.478 & 0.471 \\
    PopEnt strategy     & 0.504 & 0.445 & 0.489 & 0.456 & 0.473 \\
    PopError strategy   & 0.492 & 0.456 & 0.456 & \textbf{0.424} & 0.457 \\
    PopVar strategy     & 0.494 & 0.484 & 0.478 & 0.438 & 0.473 \\
    \bottomrule
  \end{tabular}
\end{table}

\begin{table*}[htbp]
\centering
\small
\caption{One-tailed $p$-values for two-sample Student-$t$ tests on mean RMSE
(row strategy vs.\ column strategy; $H_0\!:\mu_\text{row}=\mu_\text{col}$,
$H_1\!:\mu_\text{row}<\mu_\text{col}$).}
\label{tab:t_test_rmse}

\newcommand{\pval}[2]{#1\textsuperscript{#2}}

\setlength{\tabcolsep}{3pt}
\begin{tabular}{lcccccccccccc}
\toprule
\textbf{Strategies} & 1. & 2. & 3. & 4. & 5. & 6. & 7. & 8. & 9. & 10. & 11. & 12.\\
\midrule
1.\ Double DQN          & --   & 0.523 & 0.447 & \pval{0.022}{**} & \pval{0.004}{***} & \pval{0.005}{***} & \pval{0.002}{***} & \pval{0.006}{***} & \pval{0.001}{***} & \pval{0.024}{**} & \pval{0.072}{*}  & \pval{0.017}{**} \\
2.\ Dueling DQN         & 0.477 & --   & 0.445 & \pval{0.019}{**} & \pval{0.003}{***} & \pval{0.004}{***} & \pval{0.001}{***} & \pval{0.004}{***} & \pval{0.002}{***} & \pval{0.020}{**} & \pval{0.072}{*}  & \pval{0.014}{**} \\
3.\ Standard DQN         & 0.553 & 0.555 & --   & \pval{0.022}{**} & \pval{0.004}{***} & \pval{0.005}{***} & \pval{0.002}{***} & \pval{0.006}{***} & \pval{0.001}{***} & \pval{0.024}{**} & \pval{0.076}{*}  & \pval{0.017}{**} \\
4.\ Popularity strategy & 0.978 & 0.981 & 0.978 & --               & \pval{0.006}{***} & \pval{0.009}{***} & \pval{0.004}{***} & \pval{0.015}{**} & 0.450            & 0.402            & 0.726            & 0.396 \\
5.\ Gini strategy       & 0.996 & 0.997 & 0.996 & 0.994 & --      & 0.765 & 0.838 & 0.911 & 0.991 & 0.994 & 0.996 & 0.993 \\
6.\ Entropy strategy    & 0.995 & 0.996 & 0.995 & 0.991 & 0.235   & --    & 0.597 & 0.769 & 0.988 & 0.990 & 0.995 & 0.990 \\
7.\ Error strategy      & 0.998 & 0.999 & 0.998 & 0.996 & 0.162   & 0.403 & --    & 0.721 & 0.994 & 0.995 & 0.998 & 0.995 \\
8.\ Variance strategy   & 0.994 & 0.996 & 0.994 & 0.985 & \pval{0.089}{*} & 0.231 & 0.279 & --    & 0.981 & 0.981 & 0.991 & 0.981 \\
9.\ PopGini strategy    & 0.999 & 0.998 & 0.999 & 0.550 & \pval{0.006}{***} & \pval{0.006}{***} & \pval{0.019}{**} & 0.425 & --   & 0.792 & 0.416 & 0.792 \\
10.\ PopEnt strategy    & 0.976 & 0.980 & 0.976 & 0.598 & \pval{0.006}{***} & \pval{0.007}{***} & \pval{0.020}{**} & 0.981 & 0.208 & --   & 0.629 & 0.596 \\
11.\ PopError strategy  & 0.928 & 0.928 & 0.581 & 0.379 & \pval{0.000}{***} & \pval{0.000}{***} & 0.000*** & 0.991 & 0.571 & 0.629 & --    & 0.462 \\
12.\ PopVar strategy    & 0.983 & 0.986 & 0.623 & 0.404 & \pval{0.000}{***} & \pval{0.000}{***} & 0.000*** & 0.981 & 0.608 & 0.538 & 0.462 & --   \\
\bottomrule
\end{tabular}

\vspace{0.3em}
\footnotesize
\textit{Note}: $*$ $p<0.10$, $**$ $p<0.05$, $***$ $p<0.01$.
\end{table*}

The experimental results, presented in Table~\ref{tab:strategy_rmse}, compare the performance of various recommendation strategies across different numbers of shown items (10, 25, 50, and 100) using Root Mean Square Error (RMSE) as the evaluation metric. The strategies include three Deep Q-Network (DQN) variants—standard DQN, Double DQN, and Dueling DQN—and a range of non-personalized baselines, such as Popularity and its variants (e.g., PopGini, PopError).

The DQN-based methods consistently outperform most non-personalized baselines in terms of mean RMSE. Dueling DQN achieves the lowest mean RMSE of 0.429, closely followed by Double DQN and standard DQN, both at 0.430. This suggests that the architectural enhancements in Dueling DQN and Double DQN provide slight but meaningful improvements over the standard DQN. 

Dueling DQN achieves the lowest RMSE (0.408), outperforming Double DQN (0.425) and standard DQN (0.431). Non-personalized baselines, such as Popularity (0.497) and Gini (0.672), exhibit higher errors, suggesting that DQN-based methods excel in early-stage preference elicitation due to their adaptive learning capabilities. Double DQN takes the lead with an RMSE of 0.426, followed by standard DQN (0.428) and Dueling DQN (0.436). Popularity improves to 0.471, but remains less effective than the DQN variants.
Standard DQN achieves the lowest RMSE (0.421), with Dueling DQN (0.432) and Double DQN (0.439) close behind. Popularity baselines continue to improve, narrowing the gap with DQN methods. PopError emerges as the top performer with an RMSE of 0.424, slightly surpassing Double DQN (0.429), Dueling DQN (0.441), and standard DQN (0.442). This indicates that certain baselines can leverage extensive feedback to rival or exceed DQN-based methods in later stages. These trends suggest that DQN-based methods are particularly effective in the early to mid-range of interactions (10–50 items), while some baselines, like PopError, gain ground with more data.

Pairwise comparisons using one-tailed $t$-tests on mean RMSE (Table~\ref{tab:t_test_rmse}) confirm the statistical significance of these findings. The DQN variants significantly outperform several baselines: Double DQN vs. Popularity ($p < 0.05$), Gini ($p < 0.01$), and Entropy ($p < 0.01$). Dueling DQN and standard DQN show similar significant improvements over these baselines. However, differences among the DQN variants are not statistically significant (Double DQN vs. Dueling DQN, $p = 0.523$), indicating comparable performance. Notably, the gap between DQN methods and top baselines like PopError ($p > 0.05$) is not significant, suggesting that while DQN methods lead on average, certain heuristics remain competitive.
\section{Discussion}

The superior performance of DQN-based methods, especially Dueling DQN, highlights the advantage of reinforcement learning in cold-start recommendation systems. These methods adaptively select items to elicit user preferences, leveraging feedback to minimize RMSE effectively. The slight edge of Dueling DQN (mean RMSE 0.429) over Double DQN and standard DQN (both 0.430) likely stems from its ability to disentangle state values and action advantages, enhancing decision-making. Double DQN’s performance benefits from reduced Q-value overestimation, making it more stable than standard DQN, particularly at higher interaction counts.

These findings have practical implications for designing recommendation systems. DQN-based methods, particularly Dueling DQN, are ideal for platforms needing rapid preference elicitation for new users, such as e-commerce or streaming services. Their early-stage effectiveness (e.g., RMSE 0.408 at 10 items) can enhance user onboarding. However, for systems with longer interaction windows, integrating heuristics like PopError could provide a cost-effective alternative, given their simplicity and competitive late-stage performance.

The study’s reliance on RMSE may overlook other quality aspects, such as recommendation diversity or user satisfaction. The DQN models’ action space, potentially constrained to popular items, might limit personalization for users with niche preferences. Additionally, the offline evaluation using historical data assumes consistent user feedback, which may not hold in real-time settings.

Future research directions for recommendation systems could encompass several key areas. Alternative metrics, such as diversity and serendipity, should be integrated alongside traditional measures like RMSE to offer a more comprehensive evaluation of recommendation quality, capturing not only accuracy but also the variety and unexpectedness of suggestions \cite{stamenkovic2021choosingbestworldsdiverse}. Scalability efforts could focus on expanding the DQN's action space, potentially through embeddings or hierarchical policies, to accommodate larger item catalogs and adapt to real-world demands \cite{Zhang_Hao_Chen_Li_Chen_Sun_2019}. Finally, hybrid approaches combining the adaptability of DQN with the robustness of heuristic methods could enhance performance across both early and late stages of the recommendation process, creating a more resilient and effective system \cite{smith2017}.

We implemented the entire pipeline in Python 3.6 using the Stable Baselines library on a workstation with an Intel i7 CPU, an NVIDIA RTX 3070 Ti GPU, and 512 GB SSD storage. Each variant was trained for 2,000 episodes per recommendation list (10, 25, 50, or 100 items), requiring approximately six hours of wall-clock time. From a theoretical standpoint, the DQN update scales polynomially in the number of states and actions; by restricting our state and action spaces to the 200 most popular items, we reduce the cubic complexity to a constant factor.

In conclusion, we presented one of the first systematic explorations of advanced Q-network architectures for the cold-user recommendation problem under privacy constraints. Our results show that Double and Dueling DQNs consistently outperform traditional heuristics and provide stable improvements over standard DQN, highlighting the promise of reinforcement learning for adaptive preference elicitation. While the gains are modest, the findings establish a foundation for RL-driven recommenders that do not rely on demographic data, a key requirement in GDPR-era systems. Looking forward, extending evaluation to user-centric measures (e.g., diversity, serendipity) and scaling to larger action spaces will be crucial for bridging research insights with deployment in real-world platforms.

\newpage

\nocite{langley00}

\bibliography{example_paper}
\bibliographystyle{icml2025}

%%%%%%%%%%%%%%%%%%%%%%%%%%%%%%%%%%%%%%%%%%%%%%%%%%%%%%%%%%%%%%%%%%%%%%%%%%%%%%%
%%%%%%%%%%%%%%%%%%%%%%%%%%%%%%%%%%%%%%%%%%%%%%%%%%%%%%%%%%%%%%%%%%%%%%%%%%%%%%%
% APPENDIX
%%%%%%%%%%%%%%%%%%%%%%%%%%%%%%%%%%%%%%%%%%%%%%%%%%%%%%%%%%%%%%%%%%%%%%%%%%%%%%%
%%%%%%%%%%%%%%%%%%%%%%%%%%%%%%%%%%%%%%%%%%%%%%%%%%%%%%%%%%%%%%%%%%%%%%%%%%%%%%%
\newpage
\appendix
\onecolumn
%\section{You \emph{can} have an appendix here.}

%%%%%%%%%%%%%%%%%%%%%%%%%%%%%%%%%%%%%%%%%%%%%%%%%%%%%%%%%%%%%%%%%%%%%%%%%%%%%%%
%%%%%%%%%%%%%%%%%%%%%%%%%%%%%%%%%%%%%%%%%%%%%%%%%%%%%%%%%%%%%%%%%%%%%%%%%%%%%%%

\end{document}